\documentstyle[11pt]{article}
\date{hep-th/9801148\\ Accepted by Physics Letters B}
\title{Geometric quantization of $\bf\rm N=2, D=3$ superanyon}
\author{I.V.Gorbunov${}^1$ and S.L.Lyakhovich\\
{\small\it Department of Physics, Tomsk State University},\\
{\small\it Lenin Ave. 36, Tomsk, 634050 Russia}}
\begin{document}
\maketitle
\footnotetext[1]{e-mail: ivan@phys.tsu.tomsk.su}
\begin{abstract}
A classical model of $\rm N=2, D=3$ fractional spin superparticle
(superanyon) is presented, whose first-quantization procedure combines
the Berezin quantization for the superspin degrees of freedom and
the canonical quantization for the space-time ones. To provide the
supersymmetry for the quantised theory, certain quantum corrections are
required to the $\rm N=2$ supersymmetry generators as compared to the
Berezin procedure.  The renormalized generators are found and the first
quantised theory of $\rm N=2$ superanyon is constructed.\\[1mm]
\end{abstract}

Anyons, particles with fractional spin and statistics in (1+2)-dimensional
space-time, made theirs appearance twenty years ago \cite{ANY1}. In the
middle 80's, the anyon concept was applied to the explanation of the
fractional quantum Hall effect \cite{Hall} and to the high-$T_c$
superconductivity models \cite{Wilczekbook}.

From the group-theoretical viewpoint, a possibility of fractional spin
emerges from the ordinary classification of the
Poincar\'e group irreps. Spin is not quantised in $\bf\rm D=1+2$ because
the little group of the massive irrep $\rm SO(2)$ is an abelian group.
Fractional spin describes an appropriate representation of the universal
covering group $\overline{\rm ISO^\uparrow(1,2)}$.

The investigations of anyons in the field-theory \cite{FF,FF1} are
supplemented by the study of the mechanical models and the corresponding first
quantised theories \cite{particles,interaction,GKL1}. One of the
related problems is to realize the one-particle wave equations of anyon
and the corresponding action functionals \cite{particles} in the
form to be convenient for the quantum field theory. Another
interesting problem is to construct a consistent interaction of anyons to
external electromagnetic or gravitational fields~\cite{interaction,GKL1}.

The study of the spinning particle models \cite{Fryd} has a long
manifold history.  For a certain period these models served as test examples
of application of the modern quantization methods. In particular, the
spinning particles are treated as elementary systems within the
Kostant-Souriau-Kirillov (KSK) quantization method \cite{KSK,Wood}. Roughly
speaking, the later is based on the observation that physical phase space of
any elementary system is isomorphic to a coadjoint orbit of the symmetry
group. In the framework of the KSK construction the symplectic action of the
symmetry group (classical mechanics) lifts to a unitary irreducible
representation of the group in a space of functions on the classical
symplectic manifold (quantum mechanics).  More perfect results give the
Berezin quantization method \cite{Ber,PerMar}, which is powerful for the
K\"ahler homogeneous spaces. In this case the one-to-one correspondence can
be established between the phase-space functions (covariant Berezin symbols)
and the linear operators in a Hilbert space, being realized by
holomorphic functions on the classical manifold. Moreover, the
multiplication of the operators induces a noncommutative binary
$\ast$-operation for the covariant symbols, that provides the correspondence
principle for the observables \cite{Ber}.

The classical mechanics of $\rm D=3$ spinning particle and superparticle
has some peculiar features, which have no parallel in higher dimensions.
In particular, one can conceive that the anyon (super)particle lives in
the phase space of special (super)geometry, being a direct product
of the cotangent bundle of Minkowski space $T^\ast({\rm R}^{1,2})$
(phase space of the spinless particle) to the curved inner symplectic
(super)space, which provides the 1+2 particle with nontrivial (super)spin and
corresponding degrees of freedom. Moreover, the inner (super)manifold is
endowed with a structure of a K\"ahler homogeneous (super)space. The
structure found for anyon particle and $\rm N=1$ superparticle in Ref.\
\cite{GKL2} provides a convenient tool for first quantization of
the (super)anyon. One can combine the canonical quantization in $T^\ast({\rm
R}^{1,2})$ and the geometric quantization of the superspin degrees of freedom
\cite{GKL2}. First class constraints of the classical mechanics are converted
into one-particle wave equations of (super)anyon according to the Dirac
quantization prescription. The first quantization procedure \cite{GKL2}
gives results in agreement with the known description of the
fractional (super)spin by the use of the unitary representations of the
discrete series of $\rm su(1,1)\cong so(1,2)$ algebra \cite{FF1}
and of the unitary irreps of $\rm osp(2|2)$ superalgebra and the deformed
Heisenberg algebra \cite{Vas}.

In this letter, we suggest $\rm N=2$ superextended anyon model, generalizing
the respective $\rm N=1$ one \cite{GKL2}.
We construct the embedding of the maximal coadjoint orbit of $\rm N=2$
Poincar\'e supergroup into the phase space $T^\ast({\rm R}^{1,2})\times {\cal
L}^{1|2}$, where the inner supermanifold ${\cal L}^{1|2}\cong \rm
OSp(2|2)/U(1)\times U(1)$, to be associated to the particle superspin, is a
K\"ahler $\rm OSp(2|2)$\--ho\-mo\-ge\-neous superspace, a typical orbit of
the coadjoint representation of $\rm OSp(2|2)$. The supergeometry underlying
${\cal L}^{1|2}$ and the geometric quntization is studied in
\cite{Balant,Grad}. The first quantization procedure has some new features
for $\rm N=2$ superparticle if compared to the $\rm N=1$ case \cite{GKL2}.
Surprisingly, certain quantum corrections are required to the $\rm N=2$
supercharge's operators, being originally constructed from the respective
classical values by an ordinary correspondence rule. It should be noted that
there is no general prescription to make such corrections in the framework
of Berezin quantization. Nevertheless, these corrections, being crucial for a
compatibility of conventional quantum theory, can be exactly computed in
this case.

Turn to explicit constructions. As is known
\cite{BMS,FF1,particles,interaction}, the dynamics of the $\rm D=3$
spinning particle of mass $m$ and spin $s$ can be realized in the six
dimensional phase space with symplectic two-form
\footnote{We use Latin
letters $a,b,c,\dots$ to denote vector indices and Greek letters
$\alpha,\beta,\gamma\dots$ for spinor ones; the space--time metric
is $\eta_{ab}={\rm diag}(-,+,+)$, the antisymmetric Levi-Civita tensor
$\epsilon^{abc}$ is normalized by the condition $\epsilon^{012}=-1$;
the spinor indices are raised and lowered with the
use  of the spinor metric by the rule
$\psi_{\alpha}=\epsilon_{\alpha\beta}\psi^{\beta}, \psi^{\alpha}=
\epsilon^{\alpha\beta}\psi_{\beta}$, $\epsilon^{\alpha\beta}=-\epsilon^{\beta
\alpha}= -\epsilon_{\alpha\beta}\,,\epsilon^{01}=-1$.}
\begin{equation}
\Omega_s=-dx^a\wedge dp_a+s\Omega_{\rm m}\qquad\qquad
\Omega_{\rm m}=\frac{1}{2}\frac{\epsilon^{abc}p_a dp_b\wedge
dp_c}{(-p^2)^{3/2}}
\label{1}
\end{equation}
and is generated by the only constraint
\begin{equation}
p^2+m^2=0\,.                    \label{2}
\end{equation}
The closed form $\Omega_{\rm m}$ is called the Dirac monopole two-form.
This formulation of the classical dynamics of anyon (known
as {\it canonical\/}) is well suited for the introducing of external
fields~\cite{interaction}, but it is inconvenient for first quantization,
because in quantum theory the {\it covariant\/} realization of the nonlinear
Poisson structure (\ref{1}) remains unsolved problem. To avoid this problem,
one may equivalently reformulate the model in an extended phase space
\cite{particles,GKL1}.

The following reformulation is suitable for constructing the superextension
of the model. Let us observe that $\Omega_{\rm m}$ (\ref{1}) is a
K\"ahler two-form in a Lobachevsky plane $\cal L$ realized as a
mass hyperboloid~(\ref{2}). Really, consider the stereographic mapping of the
mass hyperboloid onto an open unit disc in complex plane
\begin{equation}
p_a\equiv mn_a\label{3}
\end{equation}
\begin{displaymath}
n_a\equiv -\left(\frac{1+z\bar{z}}{1-z\bar{z}},
\frac{z+\bar{z}}{1-z\bar{z}},i\frac{z-\bar{z}}{1-z\bar{z}}\right)\qquad
|z|<1\qquad n^2\equiv-1\;.
\end{displaymath}
Thus we are arriving at the Poincar\'e realization of $\cal L$. Using the
complex coordinate $z$, we have $\Omega_{\rm m}=1/2\epsilon^{abc}n_a
dn_b\wedge dn_c=-2i(1-z\bar{z})^{-2}dz\wedge d\bar{z}$. One can reformulate
now the canonical model (\ref{1}) in terms of the eight dimensional phase
space ${\cal M}^8=T^\ast({\rm R}^{1,2})\times{\cal L}$ with the symplectic
two-form
\begin{equation}
\Omega_s=-dx^a\wedge dp_a-2is\frac{dz\wedge d\bar{z}}{(1-z\bar{z})^2}\label{4}
\end{equation}
and three constraints (\ref{3}), two of which are of
the second class and one of the first class.  These constraints are
equivalent to the following two first class constraints
\begin{equation}
p^2+m^2=0\qquad\qquad (p,n)+m=0\,, \label{5}
\end{equation}
that should mean the identical conservation of the particle
mass and spin in ${\cal M}^8$. This constrained Hamiltonian theory
could be derived directly from the first-order covariant Lagrangian
\begin{equation}
L=m(\dot{x},n)+is\frac{\bar{z}\dot{z}-z\dot{\bar{z}}}
{1-z\bar{z}}\;.\label{6}
\end{equation}

The geometry of the symplectic manifold ${\cal M}^8$ is well
adapted for the first quantization. We can canonically quantize the Poisson
bracket in $T^\ast({\rm R}^{1,2})$ and to apply the Berezin quantization
method \cite{Ber} in the Lobachevsky plane \cite{Ber,PerMar}. The constraints
(\ref{5}) are converted into the anyon wave equations at the quantum level.
The quantization of the model (\ref{6}) is described in~Ref.~\cite{GKL2}.

Introduce the following $\rm N=2$ superextension of the Lagrangian
(\ref{6}):
\begin{equation}
L=m(\dot{x},n)-im(n_{\alpha\beta}\theta^\alpha\dot{\theta}^\beta+
n_{\alpha\beta}\chi^\alpha\dot{\chi}^\beta)+mb(\theta_\alpha\dot{\chi}^\alpha
-\chi_\alpha\dot{\theta}^\alpha)-mb\theta^\alpha n_{\alpha\gamma}
\dot{n}^\gamma{}_\beta\chi^\beta+is\frac{\bar{z}\dot{z}-z\dot{\bar{z}}}
{1-z\bar{z}}\;,\label{7}
\end{equation}
where $n_{\alpha\beta}=(n^a\sigma_a)_{\alpha\beta}$, the explicit form of
$\sigma$-matrices
\begin{displaymath}
(\sigma_0)_{\alpha\beta}=
\left( \begin{array}{cc}
            0 & 1 \\
            1 & 0
       \end{array}\right)\quad
(\sigma_1)_{\alpha\beta}=
\left( \begin{array}{cc}
            1 & 0 \\
            0 & 1
       \end{array}\right)\quad
(\sigma_2)_{\alpha\beta}=
\left( \begin{array}{cc}
           -i  & 0 \\
            0 & i
       \end{array}\right)
\end{displaymath}
is compatible with the $\rm SU(1,1)$ spinor formalism;
$\theta^\alpha,\chi^\alpha$ are $\rm D=3$ real odd Grassmann spinors.
As is shown below, the real parameter $b$ is related to the
superparticle's central charge. Besides the Poincar\'e group, the model
(\ref{7}) is invariant under the supertranslations of the form
\begin{equation}
\begin{array}{llll} \displaystyle
\delta_\epsilon
x^a=i(\sigma^a)_{\alpha\beta}\epsilon^\alpha\theta^\beta+ ib\epsilon^{abc}
n_b (\sigma_c)_{\alpha\beta}\epsilon^\alpha\chi^\beta
-bn^a\epsilon^\alpha\chi_\alpha &
\displaystyle
\delta_\epsilon \theta^\alpha=\epsilon^\alpha &
\displaystyle
\delta_\epsilon \chi^\alpha=0 &
\displaystyle
\delta_\epsilon z=0 \\
\displaystyle
\delta_\eta x^a=i(\sigma^a)_{\alpha\beta}\eta^\alpha\chi^\beta-
ib\epsilon^{abc} n_b (\sigma_c)_{\alpha\beta}\eta^\alpha\theta^\beta
+bn^a\eta^\alpha\theta_\alpha &
\displaystyle
\delta_\eta \theta^\alpha=0 &
\displaystyle
\delta_\eta \chi^\alpha=\eta^\alpha &
\displaystyle
\delta_\eta z=0\,.
\end{array}
\label{8}
\end{equation}
The corresponding supercharges $Q_\alpha^I,I=1,2$ generate the Poisson
brackets
\begin{equation}
\{{\cal Q}^I_\alpha\;,\;{\cal Q}^J_\beta\}=
-2i\delta^{IJ}p_{\alpha\beta}-2\epsilon^{IJ}
\epsilon_{\alpha\beta}{\cal Z}\; \qquad {\cal Z}=-b(p,n)\approx mb\,,\label{9}
\end{equation}
where ${\cal Z}$ is a central charge and $\approx$ means a weak equality
(modulo constraints). Because of the
Bogomol'nyi-Prassad-Sommerfield bound $m\geq |{\cal Z}|$ (see, for instance,
\cite{sohn}) one may take here $|b|\leq1$. Moreover, one can easy verify
that in the BPS limit, when $|b|=1$, half of the odd degrees of freedom drops
out of the Lagrangian (\ref{7}) and the model reduces to the one of $\rm N=1$
fractional spin superparticle considered in Ref. \cite{GKL2}. This
case will not discussed again here and we assume $|b|<1$ below.

In the terms of symplectic geometry, the Hamiltonian dynamics of the
superparticle model~(\ref{7}) is realized in $(8|4)$-dimensional
supermanifold ${\cal M}^{8|4}$ of a special structure,
${\cal M}^{8|4}\cong T^\ast({\rm R}^{1,2})\times {\cal L}^{1|2}$, where
${\cal L}^{1|2}$ is an inner supermanifold of real dimension $(2|4)$.
This means that the symplectic two-form on ${\cal M}^{8|4}$ reads
\begin{equation}
\Omega^{\rm SUSY}_s=-dx^a\wedge dp_a+s\Omega_{{\cal L}^{1|2}}\,,\label{10}
\end{equation}
where $\Omega_{{\cal L}^{1|2}}$ does not depend on the space-time coordinates
and momenta. $\rm N=2$ Poincar\'e supersymmetry leaves invariant the
constraint surface (\ref{5}) in ${\cal M}^{8|4}$. The Hamiltonian
representation of the Poincar\'e superalgebra, being described below,
shows that the constraints (\ref{5}) provide the identical conservation laws
for the superparticle mass and superspin. These conserved values
coincide respectively with constants $m$ and $s$ entering the original
Lagrangian (\ref{7}).

In the nonsuperextended model on ${\cal M}^8$, a K\"ahler geometry of the
inner manifold $\cal L$ makes possible to apply the Berezin quantization
method for the construction of the first quantised theory of anyon. Let us
study a supergeometry underlying ${\cal L}^{1|2}$ and clarify its
relationship to $\rm N=2$ Poincar\'e superalgebra.

The supermanifold ${\cal L}^{1|2}$ is considered in Refs.\
\cite{Balant,Grad} as a typical coadjoint orbit of the $\rm OSp(2|2)$
supergroup, ${\cal L}^{1|2}\cong \rm OSp(2|2)/U(1)\times U(1)$. Moreover,
${\cal L}^{1|2}$ is shown to be a K\"ahler homogeneous
$\rm OSp(2|2)$-superspace of complex dimension $(1|2)$, a~${\rm N=2}$
superextension of the Lobachevsky plane. ${\cal L}^{1|2}$ is called as
$N=2$ {\it superunit disc.\/} In holomorphic coordinates $z,\theta,\chi$ in
${\cal L}^{1|2}$, the K\"ahler superpotential reads \cite{Grad}
\begin{equation}
\Phi=-2\ln(1-z\bar{z})-(1+b)\frac{\theta\bar\theta}{1-z\bar{z}}
-(1-b)\frac{\chi\bar\chi}{1-z\bar{z}}
+\frac{1-b^2}{2}\frac{\theta\bar\theta\chi\bar\chi}{(1-z\bar{z})^2}\;,
\label{11}
\end{equation}
and the symplectic two-superform is defined with respect to the K\"ahler
superpotential in the standard way\footnote{We use the left derivatives only.}
\begin{equation}
s\Omega_{{\cal L}^{1|2}}=\frac{is}{2}\bar\delta\delta\Phi\qquad\qquad\qquad
\delta=dz\frac{\partial}{\partial z}+d\theta\frac{\partial}{\partial\theta}+
d\chi\frac{\partial}{\partial\chi}\;.\label{12}
\end{equation}
The complex odd variables $\theta,\chi$ are in one-to-one
correspondence with the Majorana spinors $\theta^\alpha,\chi^\alpha$ used
before in Eq.\ (\ref{7})
\begin{equation}
\begin{array}{l} \displaystyle
\theta=\sqrt\frac{m}{s}(z_\alpha\theta^\alpha-iz_\alpha\chi^\alpha)
\left[1+m\frac{1-b}{4s}(\theta^\alpha\theta_\alpha+\chi^\alpha\chi_\alpha)
\right]\\ \displaystyle
\chi=\sqrt\frac{m}{s}(z_\alpha\chi^\alpha-iz_\alpha\theta^\alpha)
\left[1+m\frac{1+b}{4s}(\theta^\alpha\theta_\alpha+\chi^\alpha\chi_\alpha)
\right]\qquad z_\alpha\equiv(z,-1)\;.
\end{array}
\label{13}
\end{equation}

Our main interest here is in the supergroup of the superholomorphic
canonical transformations on ${\cal L}^{1|2}$. We have found that this
supergroup, denoted $\rm SU(1,1|2)$, is wider than $\rm OSp(2|2)$~is. We
consider here the corresponding superalgebra $\rm su(1,1|2)$. The even part
$\rm su(1,1|2)_0={\rm span}\{J_a,$ $ P_I, P_4, Z; I=1,2,3\}$ of $\rm
su(1,1|2)$ is a direct sum of the Lorentz algebra $\rm su(1,1)$, the isotopic
algebra $\rm u(2)$ and the one-dimensional centre of the superalgebra,
whereas the odd part $\rm su(1,1|2)_1$ $={\rm span}\{E^\alpha,F^\alpha,
G^\alpha,H^\alpha\}$ is an eight dimensional module of the even part. The
Hamiltonian supergenerators read
\begin{displaymath}
J_a=-sn_a\left(1-\frac{1+b}{2}\frac{\theta\bar\theta}{1-z\bar{z}}
-\frac{1-b}{2}\frac{\chi\bar\chi}{1-z\bar{z}}
+\frac{1-b^2}{2}\frac{\theta\bar\theta\chi\bar\chi}{(1-z\bar{z})^2}
\right) \qquad Z=s
\end{displaymath}
\begin{equation}
\begin{array}{l}
\begin{array}{ll} \displaystyle
P_1=s\frac{\sqrt{1-b^2}}{2}\frac{\theta\bar\chi-\bar\theta\chi}{1-z\bar{z}}&
\displaystyle
P_3=s\left(\frac{1+b}{2}\frac{\theta\bar\theta}{1-z\bar{z}}-
\frac{1-b}{2}\frac{\chi\bar\chi}{1-z\bar{z}}\right)\\ \displaystyle
P_2=is\frac{\sqrt{1-b^2}}{2}\frac{\theta\bar\chi+\bar\theta\chi}{1-z\bar{z}}&
\displaystyle
P_4=s\left(\frac{1+b}{2}\frac{\theta\bar\theta}{1-z\bar{z}}
+\frac{1-b}{2}\frac{\chi\bar\chi}{1-z\bar{z}}
-\frac{1-b^2}{2}\frac{\theta\bar\theta\chi\bar\chi}{(1-z\bar{z})^2}\right)
\end{array}\\
\begin{array}{ll}
\displaystyle
E^\alpha=s\sqrt{1+b}
\left(\frac{z^\alpha\bar\theta-\bar z^\alpha\theta}{1-z\bar z}\right)
\left(1-\frac{1-b}{2}
\frac{\chi\bar\chi}{1-z\bar{z}}\right) & \displaystyle
F^\alpha=in^\alpha{}_\beta E^\beta \\
\displaystyle
G^\alpha=s\sqrt{1-b}
\left(\frac{z^\alpha\bar\chi-\bar z^\alpha\chi}{1-z\bar z}\right)
\left(1-\frac{1+b}{2}
\frac{\theta\bar\theta}{1-z\bar{z}}\right) & \displaystyle
H^\alpha=in^\alpha{}_\beta G^\beta\,,
\end{array}
\end{array}
\label{14}
\end{equation}
where $z^\alpha\equiv(1,z)\,,\,\bar z^\alpha\equiv(\bar z,1)$,
and form $\rm su(1,1|2)$ superalgebra with respect to the graded Poisson
bracket on $\rm N=2$ superunit disc
$\{\ ,\ \}_{{\cal L}^{1|2}}\equiv\{\ ,\ \}$:
\begin{displaymath}
\begin{array}{l}
\begin{array}{lll}
\{J_a,J_b\}=\epsilon_{abc}J^c &
\displaystyle \{J_a,E^\alpha\}=\frac{i}{2}(\sigma_a)^\alpha{}_\beta E^\beta &
\displaystyle \{J_a,F^\alpha\}=\frac{i}{2}(\sigma_a)^\alpha{}_\beta F^\beta
\\[2mm]
\{P_I,P_J\}=-\epsilon_{IJK}P^K &
\displaystyle \{J_a,G^\alpha\}=\frac{i}{2}(\sigma_a)^\alpha{}_\beta G^\beta &
\displaystyle \{J_a,H^\alpha\}=\frac{i}{2}(\sigma_a)^\alpha{}_\beta H^\beta
\end{array}
\\[3mm]
\begin{array}{llll}
\displaystyle \{E^\alpha,P_1\}=\frac12 H^\alpha&
\displaystyle \{E^\alpha,P_2\}=-\frac12 G^\alpha&
\displaystyle \{E^\alpha,P_3\}=-\frac12 F^\alpha&
\displaystyle \{E^\alpha,P_4\}=-\frac12 F^\alpha\\[2mm]
\displaystyle \{F^\alpha,P_1\}=-\frac12 G^\alpha&
\displaystyle \{F^\alpha,P_2\}=-\frac12 H^\alpha&
\displaystyle \{F^\alpha,P_3\}=\frac12 E^\alpha&
\displaystyle \{F^\alpha,P_4\}=\frac12 E^\alpha\\[2mm]
\displaystyle \{G^\alpha,P_1\}=\frac12 F^\alpha&
\displaystyle \{G^\alpha,P_2\}=\frac12 E^\alpha&
\displaystyle \{G^\alpha,P_3\}=\frac12 H^\alpha&
\displaystyle \{G^\alpha,P_4\}=-\frac12 H^\alpha\qquad\ \ (15)\\[2mm]
\displaystyle \{H^\alpha,P_1\}=-\frac12 E^\alpha&
\displaystyle \{H^\alpha,P_2\}=\frac12 F^\alpha&
\displaystyle \{H^\alpha,P_3\}=-\frac12 G^\alpha&
\displaystyle \{H^\alpha,P_4\}=\frac12 G^\alpha\\[3mm]
\end{array}
\\
\begin{array}{lll}
\displaystyle \{E^\alpha,F^\beta\}=\epsilon^{\alpha\beta}(Z-P_3) &
\displaystyle \{E^\alpha,G^\beta\}=-\epsilon^{\alpha\beta}P_2 &
\displaystyle \{E^\alpha,H^\beta\}=\epsilon^{\alpha\beta}P_1 \\[2mm]
\displaystyle \{G^\alpha,H^\beta\}=\epsilon^{\alpha\beta}(Z+P_3) &
\displaystyle \{F^\alpha,H^\beta\}=-\epsilon^{\alpha\beta}P_2 &
\displaystyle \{F^\alpha,G^\beta\}=\epsilon^{\alpha\beta}P_1\\[1.5mm]
\end{array}\\ \ \,
\{E^\alpha,E^\beta\}=\{F^\alpha,F^\beta\}=
\{G^\alpha,G^\beta\}=\{H^\alpha,H^\beta\}=i(\sigma_a)^{\alpha\beta}J^a
\\[1.5mm]\ \,
\{J_a,P_I\}=0\quad \{P_I,P_4\}=0\quad \{J_a,P_4\}=0\quad
\{Z,{\rm anything}\}=0\,.
\end{array}
\label{su112}
\end{displaymath}
\addtocounter{equation}{1}%
In particular, $\rm osp(2|2)$ subsuperalgebra is generated by
$J_a,B,\sqrt{ms}V^\alpha,\sqrt{ms}W^\alpha$, where $V^\alpha, W^\alpha$
are defined below in Eqs.\ (\ref{15a}) and $B=P_3-bZ$. These $\rm osp(2|2)$
supergenerators were evaluated in Ref.\ \cite{Grad} in the framework
of the theory of the supercoherent states. We conceive here that $\rm N=2$
superunit disc is not only the typical coadjoint orbit of the $\rm OSp(2|2)$
supergroup, ${\cal L}^{1|2}\cong \rm OSp(2|2)/U(1)\times U(1)$, but it can
be treated as an atypical orbit of the supergroup $\rm SU(1,1|2)$ as well,
${\cal L}^{1|2}\cong \rm SU(1,1|2)/U(2|2)\times U(1)$.

The supersymplectic action of the $\rm N=2$ Poincar\'e
superalgebra in ${\cal M}^{8|4}$ is generated by the following combinations
of the space-time variables and the ``inner'' $\rm su(1,1|2)$-generators:
\begin{equation}
\begin{array}{l}\displaystyle
{\cal J}_a=\epsilon_{abc}x^bp^c+J_a\qquad\qquad {\cal P}_a=p_a\qquad\qquad
{\cal Z}=mb\\
\displaystyle
{\cal Q}^1_\alpha=(ip_{\alpha\beta}W^\beta+m\tilde{W}_\alpha)
[1+{\rm q}^{cl}(bP_3-\sqrt{1-b^2}\,P_2-P_4)] \\ \displaystyle
{\cal Q}^2_\alpha=(ip_{\alpha\beta}V^\beta+m\tilde{V}_\alpha)
[1+{\rm q}^{cl}(bP_3+\sqrt{1-b^2}\,P_2-P_4)]\,,
\end{array}
\label{15}
\end{equation}
where
\begin{equation}
\begin{array}{ll}
\displaystyle
W^\alpha=\frac{1}{2\sqrt{ms}}(\sqrt{1+b}\,E^\alpha+\sqrt{1-b}\,H^\alpha) &
\displaystyle\qquad\tilde W^\alpha=
\frac{1}{2\sqrt{ms}}(\sqrt{1+b}\,F^\alpha-\sqrt{1-b}\,G^\alpha)\\
\displaystyle
V^\alpha=\frac{1}{2\sqrt{ms}}(\sqrt{1+b}\,F^\alpha+\sqrt{1-b}\,G^\alpha) &
\displaystyle\qquad\tilde V^\alpha=
\frac{1}{2\sqrt{ms}}(\sqrt{1+b}\,H^\alpha-\sqrt{1-b}\,E^\alpha)
\end{array}
\label{15a}
\end{equation}
and
\begin{equation}
{\rm q}^{cl}=\frac{1}{4s}\label{16}
\end{equation}
is a parameter, which should be renormalized later to provide the
supersymmetry in the quantum theory.

Relations (\ref{15}) assume that the problem of operator realization of
$\rm N=2$ Poincar\'e supersymmetry can be solved in the quantum theory if
appropriate realization is constructed for $\rm su(1,1|2)$ superalgebra. The
later problem may admit an elegant solution in the framework of the geometric
quantization on the K\"ahler homogeneous superspaces.
For $\rm N=2$ superunit disc the geometric quantization is constructed in
Ref.\ \cite{Grad}, where the classical symbols of $\rm osp(2|2)$ are lifted
to the operators acting in the space ${\cal O}_{s,b}$ of the antiholomorphic
functions of the form
\begin{equation}
F(\bar{z},\bar{\theta},\bar{\chi})=F_0(\bar{z}) +\sqrt{s(1+b)}\,\bar\theta
F_1(\bar{z})+\sqrt{s(1-b)}\,\bar\chi F_2(\bar{z})
+\sqrt{s(2s+1)(1-b^2)/2}\,\bar\theta\bar\chi F_3(\bar{z})\,.\label{17}
\end{equation}
We take $s>0$ below; the case of $s<0$ requires some inessential changes.
The Hamiltonian generators
(\ref{14}) of $\rm su(1,1|2)$ superalgebra may be lifted to the unitary
irreducible representation in the space ${\cal O}_{s,b}$. One gets
\begin{displaymath}
\begin{array}{l}\displaystyle
{\bf J}_a=-\bar\xi_a\bar\partial- (\bar\partial\bar\xi_a)\left(
s+\frac12\bar\theta\frac{\partial}{\partial\bar\theta}
+\frac12\bar\chi\frac{\partial}{\partial\bar\chi}\right)
\end{array}
\end{displaymath}
\begin{equation}
\begin{array}{ll}\displaystyle
{\bf P}_1=-\frac{1}{\sqrt{1-b^2}}\left(\frac{1-b}{2}\bar\chi
\frac\partial{\partial\bar\theta}
+\frac{1+b}{2}\bar\theta\frac\partial{\partial\bar\chi}\right) &\displaystyle
{\bf P}_3=\frac12\bar\theta\frac\partial{\partial\bar\theta}
-\frac12\bar\chi\frac\partial{\partial\bar\chi} \\ \displaystyle
{\bf P}_2=\frac{i}{\sqrt{1-b^2}}\left(\frac{1-b}{2}\bar\chi
\frac\partial{\partial\bar\theta}
-\frac{1+b}{2}\bar\theta\frac\partial{\partial\bar\chi}\right) &\displaystyle
{\bf P}_4=\frac12\bar\theta\frac\partial{\partial\bar\theta}
+\frac12\bar\chi\frac\partial{\partial\bar\chi}
\end{array}\label{17a}
\end{equation}
\begin{displaymath}
\begin{array}{l}\displaystyle
{\bf E}^\alpha=\frac{\sqrt{1+b}}{2}\bar\theta\left[\bar{z}^\alpha\bar\partial
+(\bar\partial\bar{z}^\alpha)\left(2s+\bar\chi\frac\partial{\partial\bar\chi}
\right)\right]
-\frac1{\sqrt{1+b}}\bar{z}^\alpha\frac\partial{\partial\bar\theta}
\\ \displaystyle
{\bf F}^\alpha=
-i\frac{\sqrt{1+b}}{2}\bar\theta\left[\bar{z}^\alpha\bar\partial
+(\bar\partial\bar{z}^\alpha)\left(2s+\bar\chi\frac\partial{\partial\bar\chi}
\right)\right]
-i\frac1{\sqrt{1+b}}\bar{z}^\alpha\frac\partial{\partial\bar\theta}
\\ \displaystyle
{\bf G}^\alpha=\frac{\sqrt{1-b}}{2}\bar\chi\left[\bar{z}^\alpha\bar\partial
+(\bar\partial\bar{z}^\alpha)\left(2s+\bar\theta\frac\partial{\partial\bar
\theta}\right)\right]
-\frac1{\sqrt{1-b}}\bar{z}^\alpha\frac\partial{\partial\bar\chi}
\\ \displaystyle
{\bf H}^\alpha=-i\frac{\sqrt{1-b}}{2}\bar\chi\left[\bar{z}^\alpha\bar\partial
+(\bar\partial\bar{z}^\alpha)\left(2s+\bar\theta\frac\partial{\partial\bar
\theta}\right)\right]
-i\frac1{\sqrt{1-b}}\bar{z}^\alpha\frac\partial{\partial\bar\chi}\,,
\end{array}
\end{displaymath}
where $\bar\partial\equiv\partial/\partial\bar{z}\,,
\bar\xi_a\equiv-1/2(2\bar{z},1+\bar{z},i(1-\bar{z}))$.
The (anti) commutation relations for these operators follow from Eqs.\ (15)
by replacing $\{\ ,\ \}\to 1/i[\ ,\ ]_\mp$ (anticommutator for
two odd operators and commutator in the rest cases).
These operators are Hermitian with respect to an invariant inner product
$\langle\cdot|\cdot\rangle^s_{{\cal L}^{1|2}}$ \cite{Balant,Grad}.
With respect to the $\rm su(1,1)$ subalgebra, the constructed representation
is decomposed into the direct sum $D^s_+\bigoplus D^{s+1/2}_+\bigoplus
D^{s+1/2}_+\bigoplus D^{s+1}_+$ of the unitary representations of discrete
series.

As the geometry of the phase space and its symmetries are clear now, we are
in position to study the quantization of the superanyon model. Consider the
space of functions of the form
$F(p,\bar{z},\bar\theta,\bar\chi)$, where $F(p,\bar{z},\bar\theta,\bar\chi)
=F_p(\bar{z},\bar\theta,\bar\chi) \in{\cal O}_{s,b}$ for each fixed momentum
$p$, and take, accounting for Eq.~(\ref{15}), the following ansatz
for the generators of the $\rm N=2$ Poincar\'e superalgebra
\begin{equation}
\begin{array}{l}\displaystyle
\hat{\cal J}_a=-i\epsilon_{abc}p^b\frac\partial{\partial p_c}
+{\bf J}_a\qquad\qquad \hat{\cal P}_a=p_a\qquad\qquad \hat{\cal Z}=mb\\
\displaystyle
\hat{\cal Q}^1_\alpha=(ip_{\alpha\beta}{\bf W}^\beta+m\tilde{\bf W}_\alpha)
[1+{\rm q}(b{\bf P}_3-\sqrt{1-b^2}\,{\bf P}_2-{\bf P}_4)] \\ \displaystyle
\hat{\cal Q}^2_\alpha=(ip_{\alpha\beta}{\bf V}^\beta+m\tilde{\bf V}_\alpha)
[1+{\rm q}(b{\bf P}_3+\sqrt{1-b^2}\,{\bf P}_2-{\bf P}_4)]\,.
\end{array}
\label{18}
\end{equation}
The operators
${\bf W}^\alpha,\tilde{\bf W}^\alpha,{\bf V}^\alpha,\tilde{\bf V}^\alpha$
are expressed as linear combinations of
${\bf E}^\alpha,\tilde{\bf F}^\alpha$, ${\bf G}^\alpha$, $\tilde{\bf
H}^\alpha$ according to relations (\ref{15a}). So, relations (\ref{18})
represent the quantum operators resulting from the classical Poincar\'e
supergroup generators by a straightforward canonical quantization.
However, examining the respective (anti)commutators, we find that
the operators (\ref{18}) do not generate a representation of $\rm N=2$
Poincar\'e superalgebra, if ${\rm q}={\rm q}^{cl}$~(\ref{16}) as in the
expressions (\ref{15}). The problem is in anticommutators of the supercharges
which have on shell the form
\begin{equation}
[\hat{\cal Q}^I_\alpha\;,\;\hat{\cal Q}^J_\beta]_+=
2\delta^{IJ}p_{\alpha\beta}-2imb\epsilon^{IJ}
\epsilon_{\alpha\beta}+{\cal O}(s^{-2})\;,
\label{19}
\end{equation}
(compare with Eq.\ (\ref{9})).
Mention that the corrections ${\cal O}(s^{-2})$, which appear
in r.h.s.\ of the anticommutators, should be expected in advance according
to a correspondence principle. The latter follows naturally from Berezin
quantization method for K\"ahler homogeneous manifolds~\cite{Ber}, which
implies that the classical symbol of quantum commutator of two bounded
operators in a Hilbert space coincides with the Poisson bracket of respective
covariant symbols only in first order in the ``Planck constant''. In
general, the corrections in higher orders vanish only for the
generators of the Lie algebra of the symmetry group. It can be found
that the correspondence principle holds for the $\rm N=2$ superunit disc
${\cal L}^{1|2}$ too\footnote{The proof will be presented elsewhere.}, where
the parameter $s^{-1}$ serves as a ``Planck constant''. Thus,
the quantum corrections in the anticommutators (\ref{19}) originate from the
{\it nonlinearity\/} of the Poincar\'e supercharge operators (\ref{18}) in
the generators (\ref{17a}) of the ``inner'' superalgebra $\rm su(1,1|2)$.

The conventional construction of the one-particle quantum mechanics for
superanyon implies to have an exact realization of the Poincar\'e
supersymmetry, without any disclosing corrections depending on the parameters
of the model. To find the true realization, we can try, starting from
Eqs.\ (\ref{18}), (\ref{19}), to introduce a renormalized terms in the
observables (\ref{18}) for the closure of the anticommutators (\ref{19}).
However, we don't have any general reasons, which may ensure the consistency
of the renormalization procedure; a structure of possible higher
order corrections to~(\ref{18}) is unclear also.
Surprisingly, exact corrections may be found in the simplest ansatz for the
quantum observables. Namely, we find that the closure of the Poincar\'e
superalgebra is achieved by the renormalization of the only parameter
${\rm q}$ entering the expressions~(\ref{18}) of the supercharges.

It is examined by a direct calculation that the generators (\ref{18}) with
renormalized value of $\rm q$
\begin{equation}
{\rm q}^{quant}=1-\sqrt{1-\frac{1}{2s+1}}={\rm q}^{cl}+{\cal
O}(s^{-2})\;.  \label{20}
\end{equation}
form the closed Poincar\'e superalgebra and, thus, they are treated as true
quantum observables of N=2 superanyon.

The super Poincar\'e covariant equations for the wave function of
the $\rm N=2$ superanyon have the form
\begin{equation}
\begin{array}{ll}\displaystyle
(p^2+m^2)F^{\rm phys}(p,\bar{z},\bar\theta,\bar\chi)=0\\ \displaystyle
[(p,{\bf J})-m{\bf P}_4-ms]F^{\rm phys}(p,\bar{z},\bar\theta,\bar\chi)=0 \;.
\end{array}
\label{20a},
\end{equation}
that appears when the constraint operators are imposed on the physical states
$F^{\rm phys}$. It should be recognized that the
$\rm N=2$ Poincar\'e supersymmetry is realized on shell only.
The last remarkable step is that the space of solutions of the
wave equations is endowed with the structure of Hilbert space, where the
operators (\ref{18}) form a {\it unitary} representation. The respective
inner product
\begin{equation}
\langle F^{\rm phys},G^{\rm phys}\rangle=\int \frac{{\rm d}\vec{p}}{p^0}
\langle F^{\rm phys}|G^{\rm phys} \rangle^s_{{\cal L}^{1|2}}\qquad
p^0=\sqrt{\vec{p}^{\;2}+m^2}>0  \label{21}
\end{equation}
is constructed in terms of the Poincar\'e invariant measure on the mass shell
and the $\rm su(1,1|2)$ invariant inner product in ${\cal L}^{1|2}$.

Thus, the first quantised theory of $\rm N=2$ superanyon has been constructed
in general. The theory give the supersymmetric generalization of the
well known description \cite{FF1,particles} of the fractional spin states
using the unitary representations $D^s_+$ of discrete series of
$\rm\overline{SU(1,1})$. Each component of the expansion~(\ref{17}) of the
wave function $F^{\rm phys}$ describes a particle of mass $m$ and fractional
spin $s$, $s+1/2$ (two states) or $s+1$.

In this letter we have begun with a classical model (\ref{7}), being
$\rm N=2$ superextension of the canonical model of anyon, and
arrive to the first quantised theory. The reverse way is in the following.
The space of quantum states $\{|p,\bar{z},\bar\theta,\bar\chi\rangle\}$
(which are solutions of Eq.~(\ref{20a})) is labeled by the points of the
surface of constraints (\ref{5}) of the classical phase superspace
$T^\ast({\rm R}^{1,2})\times {\cal L}^{1|2}$, similar to geometric
quantization method. Moreover, we have
\begin{displaymath}
\begin{array}{ll}\displaystyle
\frac{\langle p,\bar{z},\bar\theta,\bar\chi|
\hat{\cal J}_a|p,\bar{z},\bar\theta,\bar\chi\rangle}
{\langle p,\bar{z},\bar\theta,\bar\chi|p,\bar{z},\bar\theta,\bar\chi\rangle}
={\cal J}_a
&\displaystyle
\frac{\langle p,\bar{z},\bar\theta,\bar\chi|
\hat{\cal P}_a|p,\bar{z},\bar\theta,\bar\chi\rangle}
{\langle p,\bar{z},\bar\theta,
\bar\chi|p,\bar{z},\bar\theta,\bar\chi\rangle}={\cal P}_a
\end{array}
\end{displaymath}
\begin{equation}
\frac{\langle p,\bar{z},\bar\theta,\bar\chi|
\hat{\cal Q}^I_\alpha|p,\bar{z},\bar\theta,\bar\chi\rangle}
{\langle p,\bar{z},\bar\theta,\bar\chi|p,\bar{z},\bar\theta,\bar\chi\rangle}=
{\cal Q}^I_\alpha+{\cal O}(s^{-2})\,.
\end{equation}
The last correction ${\cal O}(s^{-2})$ is related certainly to the
renormalization of the parameter ${\rm q}$ in Eq.~(\ref{18}). The possibility
of this renormalization is the most intriguing result of the geometric
quantization of the superspin degrees of freedom of $\rm N=2$ superanyon.

This work is supported in part by the INTAS-RFBR Grant No.\ 95-0829. S.L.L.
is partially supported by the RFBR Grant No.\ 96-01-00482, he also
appreciates the support from DAAD.

\end{document}